# Assessing the Frequency Response Potential of Heavy-Duty Electric Vehicles with Vehicle-to-Grid Integration in the California Power System

Xiaojie Tao[a,*], Yaoyu Fan[a], Zhaoyi Ye[a], Rajit Gadh[a]

[a] Smart Grid Energy Research Center, Mechanical and Aerospace Engineering Department, University of California, Los Angeles, CA 90095, USA

*Corresponding Author: Xiaojie Tao, University of California, Los Angeles, CA 90095, USA

Email: xiaojietao@g.ucla.edu, taoxiaojie04@gmail.com

**Abstract**: The integration of heavy-duty electric vehicles (EVs) with Vehicle-to-Grid (V2G) capability can enhance primary frequency response and improve stability in power systems with high renewable penetration. This study evaluates the technical potential of heavy-duty EV fleets to support the California power grid under three practical charging strategies: immediate charging, delayed charging, and constant-minimum-power charging. We develop a simulation framework that couples aggregated frequency dynamics with battery and charger constraints, state-of-charge management, and fleet-availability profiles. Performance is assessed using standard frequency security metrics, including nadir, rate-of-change-of-frequency, overshoot, and settling time, across credible contingency scenarios and renewable generation conditions. Results indicate that both non-V2G modes and V2G-enabled operation can contribute meaningful primary response, with V2G providing the strongest and fastest support while respecting mobility and network limits. Sensitivity analyses show that the relative benefits depend on charging strategy, control parameters, and renewable output, highlighting design trade-offs between response magnitude, duration, and battery usage. Overall, heavy-duty EV fleets—when coordinated by appropriate charging and V2G controls—offer a viable resource for strengthening primary frequency control on the California grid and mitigating stability challenges associated with increasing renewable penetration.

# 1. Introduction

The rapid growth of renewable energy penetration presents new challenges for maintaining frequency security and overall grid reliability [1–4]. Renewable resources such as wind and solar, while environmentally beneficial, contribute little to system inertia because of their power-electronic interfaces, thereby increasing the vulnerability of modern power systems to frequency excursions following generation disturbances [5–7].

Meanwhile, the electrification of heavy-duty vehicles (HDVs) is accelerating worldwide as part of broader efforts to decarbonize the transportation sector [8–10]. In California, state regulations mandate a large-scale transition toward zero-emission HDVs, creating opportunities to utilize these vehicles as distributed energy resources for grid support [11]. Heavy-duty electric vehicles (HDEVs) equipped with Vehicle-to-Grid (V2G) functionality are particularly promising, as their substantial battery capacities and high-power chargers enable them to deliver primary frequency response within seconds [12,13]. However, large-scale deployment of HDV fleets also introduces challenges, including increased feeder loading, potential local congestion from simultaneous charging, and the need for coordinated control across depots [14–16].

Most prior studies on EV-based frequency response have focused on light-duty EVs or simplified charging scenarios, with limited attention to the unique operating characteristics of HDV fleets [17–19]. The differences in vehicle size, battery capacity, charging power, and duty cycles significantly affect their interaction with the grid and, consequently, their capacity to provide frequency response [20]. Furthermore, while individual charging strategies and control modes (e.g., unidirectional V1G and bidirectional V2G) have been explored, their combined impact under realistic fleet conditions remains insufficiently understood [21–24].

This study addresses these gaps by developing a simulation-based framework to quantify the frequency response potential of heavy-duty EV fleets with V2G capability in a representative California power system. Three practical charging strategies—(1) immediate, (2) delayed, and (3) constant-minimum-power—are compared under two control modes: V1G (charging only) and V2G (charging and discharging). These scenarios represent the spectrum of operational behaviors observed in HDV depots, ranging from aggressive to conservative charging management [25,26].

The objectives of this research are to (i) evaluate how charging strategies and control modes influence system frequency stability under high renewable penetration, and (ii) identify conditions under which V2G operation yields the most benefit without compromising vehicle mobility. The findings offer insights into how HDV fleets can be integrated as flexible grid resources to mitigate renewable variability while enhancing primary frequency response capability.

The remainder of this paper is organized as follows. Section 2 describes the fundamentals of frequency control and the modeling approach. Section 3 details the HDV fleet model and control framework. Section 4 presents the simulation setup for the California test system, Section 5 discusses results, and Section 6 concludes the paper with key findings and recommendations for future work.

## 2. Frequency Control

Frequency stability is a fundamental component of power system reliability [27–29]. It reflects the instantaneous balance between electrical generation and demand, with grid frequency maintained close to 60 Hz in most interconnected systems. Even small deviations from the nominal value indicate an imbalance and, if uncorrected, may escalate into large-scale instability or even system collapse [30–32].

Figure 1 illustrates a representative frequency trajectory following a generation loss. Under normal operating conditions, frequency oscillations remain minor. However, when a generation unit trips, the immediate power deficit causes a rapid frequency decline. To mitigate this, the primary frequency response is automatically triggered within seconds, utilizing the kinetic energy stored in the rotating masses of conventional generators to arrest the frequency drop. This rapid response, typically lasting several seconds, is vital for preventing further deterioration of system stability.

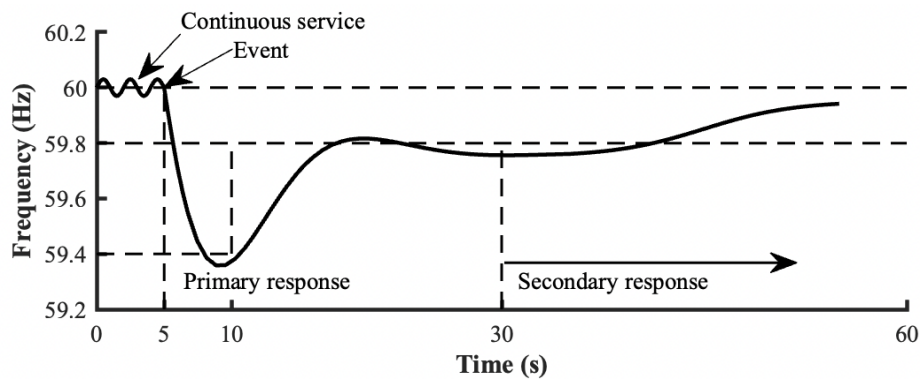

**Figure 1.** Typical power system frequency response to a sudden generation loss, showing the primary and secondary response phases. The primary response rapidly mitigates the initial frequency dip caused by the disturbance, while the secondary response gradually restores frequency to its nominal value of 60 Hz. This dynamic illustrates the importance of fast-acting resources such as Vehicle-to-Grid (V2G)-enabled heavy-duty electric vehicles that can supplement traditional generator-based responses during contingencies.

Approximately 30 seconds after the event, the secondary frequency response is activated. This process involves adjustments by automatic generation control mechanisms, which coordinate multiple generators to return system frequency to the nominal level. Together, the primary and secondary responses form the core of frequency control, ensuring stable and continuous system operation under disturbance conditions.

With the increasing penetration of renewable energy resources that lack physical rotational inertia, the inherent ability of the grid to resist frequency deviations has weakened substantially [33,34]. Consequently, alternative and faster-acting mechanisms for frequency control have become essential. In this context, the capability of electric vehicles to provide rapid, decentralized, and scalable frequency response represents a promising approach to maintaining system stability in modern low-inertia grids [35,36].

## 3. Heavy Duty Electric Vehicles

Heavy-duty electric vehicles (HDEVs) possess distinct technical characteristics that make them particularly suitable for supporting grid frequency stability [37]. Frequency deviation, analogous to voltage deviation in distribution systems, serves as a fundamental indicator of grid health [38–40]. Such deviations arise from sudden imbalances between generation and demand, often triggered by generator outages or fluctuations in renewable energy output. If not promptly corrected, frequency deviations can propagate throughout the network, leading to widespread instability or even equipment failure [41,42].

HDEVs, equipped with large-capacity batteries typically ranging from 200 kWh to more than 600 kWh, represent a transformative opportunity to mitigate frequency deviations [43,44]. Their substantial energy reserves enable them to either deliver power to the grid or absorb excess generation during frequency events. Compared with light-duty EVs, HDEVs are especially valuable because they can operate in both Vehicle-to-Grid (V2G) mode—injecting power into the grid—and unidirectional V1G mode—rapidly reducing charging demand in response to disturbances.

The fast response capability of HDEVs also makes them an ideal resource for primary frequency control. Conventional power plants may require seconds to minutes to adjust their mechanical output, whereas HDEVs, driven by advanced power-electronic converters, can respond almost instantaneously [45]. This near-instantaneous reaction is crucial in the first few seconds following a grid disturbance, when rapid power injection or reduction can arrest frequency decline and relieve the burden on slower secondary control systems.

In addition to their responsiveness, HDEVs enhance system resilience by diversifying the sources of frequency response [46,47]. Unlike centralized generating units, HDEVs can be geographically distributed across depots and charging facilities, forming a flexible, decentralized resource for grid operators. When properly coordinated, HDEV fleets not only provide dynamic frequency support but also reduce dependence on fossil-fuel-based spinning reserves, thus contributing to broader decarbonization goals.

In summary, the ability of HDEVs to store and dynamically transfer energy, combined with their fast response time and operational flexibility, positions them as a critical asset for primary frequency response. These advantages underscore the need for systematic research into HDEV control strategies and charging coordination to fully exploit their grid-support potential.

To effectively harness the frequency response capabilities of HDEVs, it is essential to consider their charging behavior and its impact on the power grid [48,49]. HDEVs typically follow scheduled operations, beginning daily shifts with a fully charged battery (100% state of charge, SoC) and returning with a partially depleted one. Ensuring sufficient recharging before the next duty cycle is critical for both fleet reliability and grid-support readiness.

This study examines three representative charging strategies—immediate, delayed, and constant-minimum-power—to evaluate their effects on grid performance. Each strategy represents a distinct approach to managing the timing and magnitude of charging, with implications for grid load distribution, operational cost, and system stability. Under the immediate charging strategy, HDEVs commence charging at full power (100 kW per vehicle in this study) immediately after completing their routes. This approach replenishes the battery to full SoC well before the next shift but often coincides with system peak demand in late afternoon and early evening hours, as illustrated in Fig. 2. The simultaneous high-power demand from multiple fleets during these periods can exacerbate grid stress, increase voltage deviations, and raise electricity costs due to higher Time-of-Use (TOU) rates.

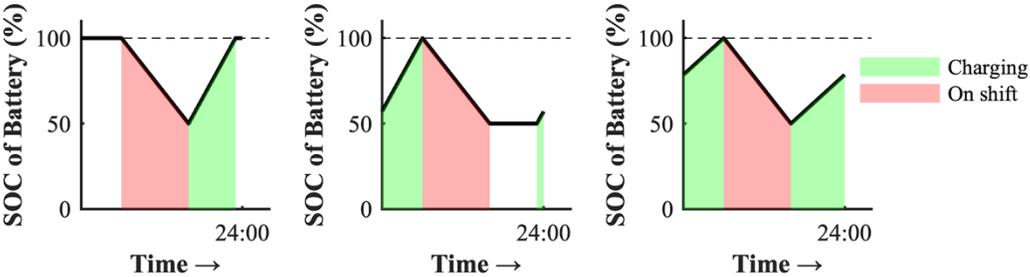

**Figure 2.** State-of-charge (SoC) trajectories of HDEV batteries under three charging strategies: immediate (left), delayed (middle), and constant-minimum-power (right).

The delayed charging strategy postpones charging to the latest feasible period before the next shift begins, ensuring full SoC just in time for operation. By avoiding peak grid demand hours, this approach reduces load stress and lowers TOU-related electricity costs. As shown in Fig. 2, the delayed strategy shifts charging activity toward early morning hours, aligning with periods of lower system demand and improving overall load distribution.

In the constant-minimum-power strategy, HDEVs initiate charging immediately after each shift but at a reduced constant power level sufficient to achieve full SoC before the next duty cycle. This approach evenly distributes charging load across the entire off-shift period, reducing the peak power per vehicle to 50 kW (Table 1) while increasing total charging duration. As depicted in Fig. 2, this strategy results in a flatter load profile and mitigates adverse impacts on grid stability during off-peak hours.

Table 1 summarizes the characteristics of the three charging strategies, including peak charging power, charging duration, and typical charging periods. While the immediate and delayed charging strategies share the same peak power of 100 kW, the constant-minimum-power approach achieves a substantially lower peak by extending the total charging time.

**Table 1.** Characteristics of the three HDEV charging strategies.

| Charging strategy | Immediate | Delayed | Constant minimum power |
|---|---|---|---|
| Peak charging power (kW) | 100 | 100 | 50 |
| Charging duration (h) | 7 | 7 | 14 |
| Charging time | 16:00 – 23:00 | 23:00 – 6:00 | 16:00 – 6:00 |

Figure 3 further illustrates the aggregate charging load profiles for HDEV fleets across California under each strategy, scaled to represent statewide deployment. Immediate charging produces a pronounced evening peak that coincides with overall grid demand. Delayed charging shifts this peak to early morning hours, while constant-minimum-power charging yields an evenly distributed load curve across the night.

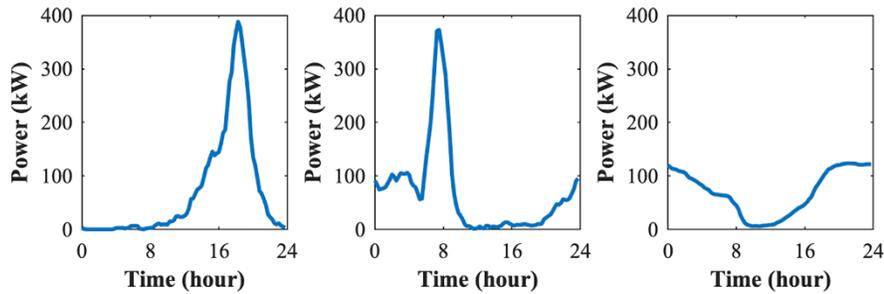

**Figure 3.** Average power demand profiles of an HDEV fleet under three charging strategies: immediate (left), delayed (middle), and constant-minimum-power (right).

These three strategies highlight the trade-offs between operational cost, fleet readiness, and grid stability. Immediate charging ensures the fastest energy replenishment but imposes the greatest stress on the power system. Delayed and constant-minimum-power charging, by contrast, provide better temporal alignment with system demand patterns, reducing network congestion and enhancing overall stability. Properly coordinated charging management allows HDEVs to deliver effective primary frequency response while minimizing their adverse impact on grid operation.

## 4. Case Study

To evaluate the potential of heavy-duty electric vehicles (HDEVs) in providing primary frequency response, a case study was performed using a Simulink-based dynamic model of the California power system, as illustrated in Fig. 4. The model incorporates key components, including conventional synchronous generators, aggregated system inertia, and HDEV charging loads represented through equivalent transfer functions. This framework enables a detailed examination of the dynamic interactions between HDEV fleets and the bulk power system during frequency disturbances.

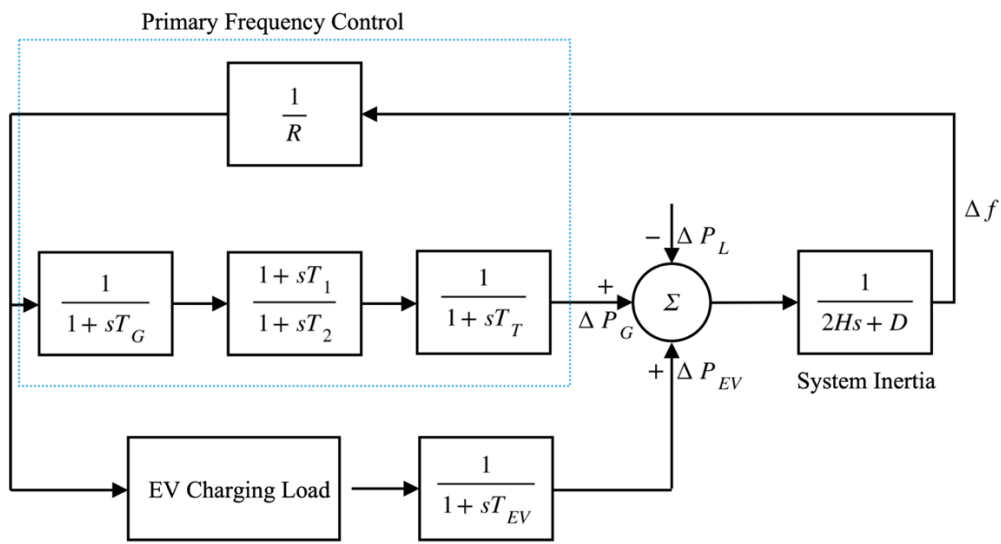

**Figure 4.** Block diagram of the primary frequency control system incorporating heavy-duty electric vehicle (HDEV) participation.

The system inertia is characterized by an effective inertia constant, Heff, calculated as the weighted average of the individual inertia constants (H) of all generation sources, where the weighting factor corresponds to each source's power contribution [36]. Table 2 summarizes the inertia constants, generation shares, and power outputs for different energy sources, highlighting the limited inertia contribution from renewable resources such as wind and solar, which lack physical rotating mass. The value of Heff was computed for a critical low-demand condition on February 28, 2021, at 20:00, when renewable generation reached its peak penetration and dispatchable generation was at its lowest level.

**Table 2.** Summary of California's generation mix and inertia contributions during the analyzed period.

| Generation source | H (s) | Generation share (%) | Generation power (MW) |
|---|---|---|---|
| Coal | 2.6 | 5.9 | 1,166 |
| Natural gas | 4.9 | 65.5 | 12,996 |
| Nuclear | 4.1 | 5.8 | 1,147 |
| Petroleum | 3.6 | 0.4 | 88 |
| Wind and solar | 0 | 4.0 | 809 |
| Hydro | 2.4 | 15.7 | 3,115 |
| Other | 0 | 2.6 | 509 |
| Effective H | 6.4 | 100 | 19,830 |

HDEVs can participate in frequency response through two distinct operational control modes:

• V1G mode (charging-only): HDEVs respond to frequency events by temporarily suspending charging, effectively reducing their instantaneous load on the grid.

• V2G mode (bidirectional): HDEVs discharge power back into the grid during disturbances, providing active power injection to counteract frequency decline.

These two response modes allow HDEVs to contribute dynamically to frequency stabilization under different grid conditions. The modeled system also includes primary droop control for synchronous generators, characterized by parameters such as the droop coefficient (R), governor time constant (TG), and turbine time constant (TT). Frequency deviations ($\Delta f$) are continuously monitored by the grid control center, and once a deviation exceeds the predefined threshold, a disturbance signal is sent to activate the HDEV control response.

In this study, the reference frequency deviation Δf corresponds to a single-step drop in system frequency caused by a generation loss of 1,800 MW. The resulting frequency trajectory reflects the combined effect of synchronous generator response, system inertia, and HDEV participation under both V1G and V2G operational modes.

## 5. Simulation Results and Discussion

Simulations were carried out to assess the influence of heavy-duty electric vehicles (HDEVs) on grid frequency stability under three charging strategies—immediate, delayed, and constant-minimum-power—and two operational modes, V1G and V2G. The test system, modeled in Simulink, represents the California power grid and includes synchronous generators with droop control, aggregated HDEV charging loads modeled as controllable power elements, and a system inertia constant derived from the state's generation mix during a low-inertia operating hour. A generation loss of 1,800 MW was introduced at t = 0 s, and a grid event signal was triggered when system frequency dropped below 59.7 Hz. Frequency response was then evaluated for both V1G and V2G modes across multiple HDEV participation levels ranging from 20 % to 100 %.

Under the immediate charging strategy, HDEVs begin charging at full rated power immediately after completing their shifts, producing a concentrated demand peak that coincides with evening grid load. The corresponding frequency responses are shown in Fig. 5. In V1G mode, increasing HDEV participation improves the frequency nadir from 59.30 Hz to 59.55 Hz as additional charging load is shed during the event. In contrast, V2G mode enables active power injection and achieves a much stronger effect, with frequency nadirs reaching up to 59.75 Hz. The dynamic recovery is also faster, with system frequency stabilizing within approximately 20 s. These results confirm that V2G operation provides both deeper and faster frequency support following a disturbance.

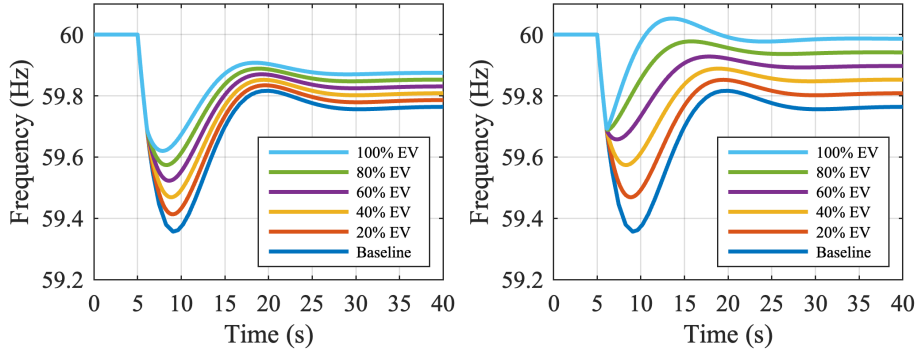

**Figure 5.** Frequency response of the grid under the immediate charging strategy with varying levels of HDEV participation in primary frequency control.

When the delayed charging strategy is applied, charging is shifted to off-peak hours, thereby reducing grid stress during critical demand periods. As illustrated in Fig. 6, the overall frequency response patterns are comparable to those of immediate charging but without aggravating peak-hour congestion. The temporal flexibility of this strategy also reduces operational costs for fleet operators due to lower Time-of-Use (TOU) electricity rates.

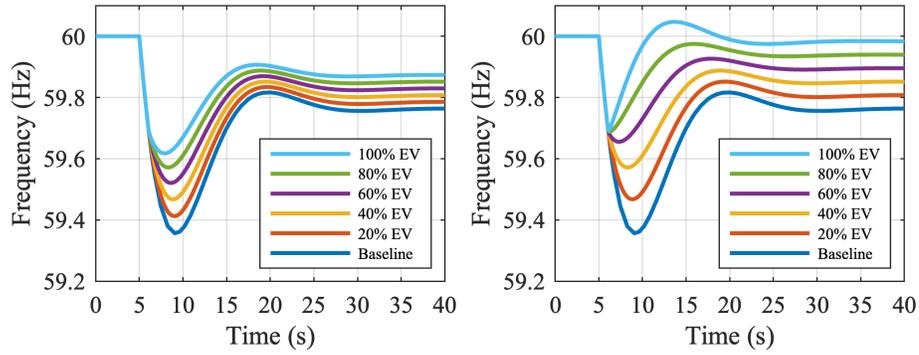

**Figure 6.** Frequency response of the grid under the delayed charging strategy with varying levels of HDEV participation in primary frequency control.

The constant-minimum-power charging strategy distributes charging uniformly over the entire off-shift period, resulting in a flatter aggregate load profile. As shown in Fig. 7, this configuration produces higher frequency nadirs during grid disturbances because of the more balanced availability of instantaneous load-shedding or power-injection capability. The smoother demand curve achieved under this strategy reduces grid stress while providing consistent support throughout the charging window.

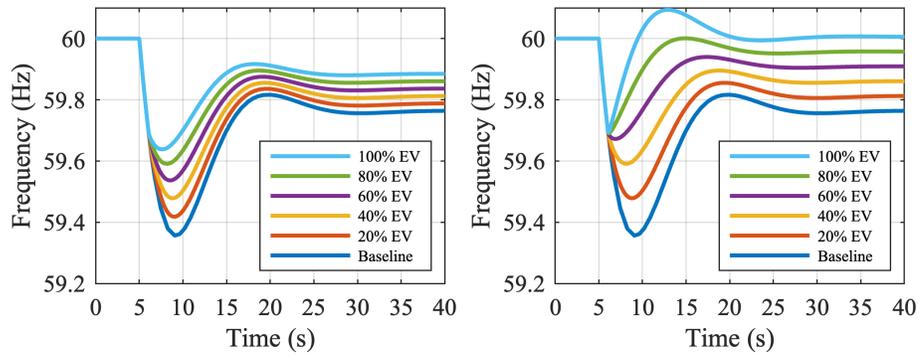

**Figure 7.** Frequency response of the grid under the constant-minimum-power charging strategy with varying levels of HDEV participation in primary frequency control.

The system frequency nadir—the lowest frequency point following a generation loss—varies according to time of day, renewable energy share, and fleet participation level. Figure 8 presents the frequency nadir values for each 15-minute interval throughout a representative day. Higher levels of HDEV participation, particularly under V2G operation, consistently yield higher nadirs, thereby mitigating the impact of generation outages. This improvement is most pronounced during periods of high renewable penetration when grid inertia is at its minimum.

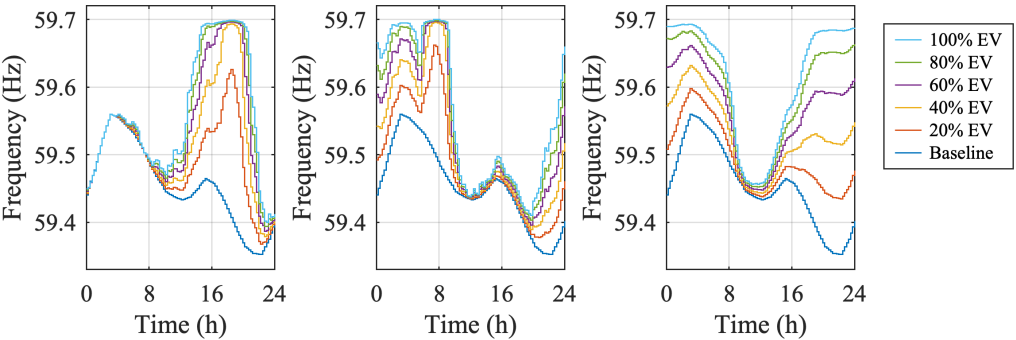

**Figure 8.** Frequency nadir after a loss of generation every 15 minutes along a day.

The simulation outcomes clearly demonstrate the superiority of V2G mode over V1G mode in maintaining frequency stability. By actively injecting power into the grid instead of merely suspending charging, V2G operation enables HDEVs to provide a stronger and faster stabilizing effect during disturbances. This capability becomes especially valuable under low-inertia conditions, when immediate and substantial power support is required to arrest frequency decline. In comparison, V1G mode offers limited benefit through load shedding alone, underscoring the enhanced effectiveness of V2G technology for dynamic grid support.

The selected charging strategy also exerts a substantial influence on overall frequency performance and system impact. Among the three strategies, the constant-minimum-power approach provides the most balanced outcome, distributing the load evenly during off-peak hours and enhancing frequency support with minimal network stress. The delayed charging strategy effectively lowers operational costs by avoiding peak demand periods, whereas the immediate charging strategy ensures full operational readiness but intensifies grid loading during evening peaks. Furthermore, the magnitude of HDEV contributions depends on dynamic system conditions such as demand variability and renewable output across the day. This variability emphasizes the need for adaptive, optimized charging coordination to maximize the value of HDEVs as a flexible resource for maintaining grid stability.

## 6. Conclusion and Future Work

This study investigated the potential of heavy-duty electric vehicles (HDEVs) to provide primary frequency response within the California power system, emphasizing their role in enhancing grid stability under high renewable penetration. Using a dynamic Simulink-based model, HDEV charging load profiles were scaled to represent statewide operations, and their interactions with the grid were analyzed under multiple charging strategies and control modes. The effective system inertia was calculated based on the state's generation mix, accounting for the reduced contribution of inverter-based renewable resources and providing a realistic foundation for evaluating system performance.

The results confirm that HDEVs constitute a promising resource for strengthening frequency stability, particularly in mitigating the effects of low system inertia. Their performance depends strongly on the adopted control mode, charging strategy, and timing of disturbances. Among all configurations, HDEVs operating in Vehicle-to-Grid (V2G) mode achieved the most significant improvement in frequency nadir and recovery speed, owing to their ability to actively inject power into the grid during contingencies. These findings highlight the importance of coordinated charging management and V2G integration to fully exploit the flexibility and responsiveness of HDEV fleets as distributed frequency-support resources.

Future work will extend this analysis by incorporating detailed distribution network constraints, stochastic fleet availability, and communication delays to further assess real-world deployment feasibility. Experimental validation using hardware-in-the-loop platforms and large-scale aggregator coordination will also be explored to demonstrate the scalability of HDEV-based frequency response solutions in next-generation low-inertia grids.

**CRediT authorship contribution statement**

**Xiaojie Tao**: Conceptualization, Methodology, Software, Validation, Formal analysis, Investigation, Writing - Original Draft, Writing - Review & Editing, Visualization. **Yaoyu Fan**: Investigation, Writing - Original Draft. **Zhaoyi Ye**: Investigation, Writing - Original Draft. **Rajit Gadh**: Supervision, Project administration, Funding acquisition, Investigation, Resources, Writing – Review & Editing.

**Declaration of competing interest**

The authors declare that they have no known competing financial interests or personal relationships that could have appeared to influence the work reported in this paper.

**Data Availability**

Data will be made available on request.

**Acknowledgements**

This work was supported by the Department of Mechanical and Aerospace Engineering, University of California, Los Angeles, under grant numbers 69763, 77739, and 45779. The authors would like to thank Dr. Kourosh Sedghi-Sigarchi for valuable discussions and constructive feedback during the preparation of this work, and Sameer Bajaj for assistance with data processing and simulation tasks.